\RequirePackage{fix-cm}
\documentclass[smallextended]{svjour3}       
\smartqed  
\usepackage{graphicx}
\usepackage[authoryear,round]{natbib}
\def\@biblabel#1{#1.}%

\usepackage{amsmath}
\usepackage{amssymb}
\usepackage{amsfonts}
\usepackage{bm}
\usepackage{caption}
\usepackage{subcaption}
\usepackage{dsfont}
\usepackage{geometry}
\usepackage[hidelinks]{hyperref}
\usepackage[inline]{enumitem}
\usepackage{rotating}
\usepackage{graphicx,xcolor}
\usepackage{float}
\usepackage{accents}
\newcommand{\1}{\mathds{1}}
\newcommand{\p}{\mathbb{P}}
\newcommand{\E}{\mathbb{E}}
\newcommand{\dd}{\mathrm{d}}

\makeatletter
\renewcommand{\makeheadbox}{} 
\makeatother

\makeatletter
\renewcommand{\@maketitle}{
	\newpage
	\null
	\vskip 2em%
	\begin{center}%
		{\Large\bfseries \@title \par}%
		\vskip 1.5em%
		{\large
			\lineskip .5em%
			\begin{tabular}[t]{c}%
				\@author
			\end{tabular}\par}%
		\vskip 1em%
	\end{center}%
	\par\vskip 1.5em%
}
\makeatother
\begin{document}
	
	\title{Can we detect treatment effect waning\\ from time-to-event data?
	}
	
	
	
	
	\author{Eni Musta      \and
		Joris Mooij\\[0.2cm]
		Korteweg-de Vries Institute for Mathematics, University of Amsterdam
	}
	
	
	\institute{Eni Musta (corresponding author) \and Joris Mooij \at
		Korteweg-de Vries Institute for Mathematics, University of Amsterdam, Amsterdam, Netherlands \\
		\email{e.musta@uva.nl}           
	}
	
	\date{Received: date / Accepted: date}

	\maketitle
	
	\begin{abstract}  
	Understanding how the causal effect of a treatment evolves over time, including the potential for waning, is important for informed decisions on treatment discontinuation or repetition. For example, waning vaccine protection influences booster dose recommendations, while cost-effectiveness analyses require accounting for long-term efficacy of treatments. However, there is no consensus on the methodology to assess and account for treatment effect waning. Even in randomized controlled trials, the common naïve comparison of hazard functions can lead to misleading causal conclusions due to inherent selection bias. Although comparing survival curves is {sometimes} recommended as a safer measure of causal effect, it only represents a cumulative effect over time and does not address treatment effect waning. We also explore recent formulations of causal hazard ratios, based on the principal stratification approach or the controlled direct effect. These causal hazard ratios cannot 
	 be identified without strong modeling assumptions, but bounds can be derived accounting for unobserved heterogeneity and one could try to use them to detect treatment effect waning.  
	However, we illustrate that an increase in causal hazard ratios towards one does not necessarily mean that the protective effect of the treatment is fading.  Furthermore, the same survival functions may correspond to both scenarios with and without waning, which shows that treatment effect waning cannot be identified from standard time-to-event data  without strong untestable
	modeling assumptions.
		\keywords{treatment efficacy\and hazard ratio \and causal inference \and randomized trials \and selection bias}
		\subclass{62G99 \and 62N99 \and 62P10} 
	\end{abstract}

	\section{Introduction}
	Waning of treatment effect refers to the gradual decline in the efficacy or benefit of a treatment over time. This phenomenon is observed in various medical and public health interventions, such as vaccines, where initial protection against disease may diminish months or years after administration, necessitating booster doses. Waning can also occur for pharmacological or other types of treatments due to biological, behavioral, or environmental factors. For instance, increased drug tolerance can contribute to reduced effectiveness over time. Understanding treatment effect waning is critical for optimizing intervention strategies, including the timing of re-treatment or discontinuation, and for accurately estimating long-term treatment benefits in cost-effectiveness analyses. However, characterizing, quantifying and interpreting  waning is complex and challenging since risk profiles change over time and a diminishing treatment effect could also result from other mechanisms, such as changes in exposure patterns. These complexities underscore the need for a better understanding of what we can or cannot conclude based on the observed data. 
	
	In the context of vaccine efficacy (VE), waning immunity has received a lot of attention recently \citep{piechotta2022waning,menegale2023evaluation,imai2023quantifying,goldberg2021waning,zhuang2022protection,clark2019efficacy,giannouchos2024waning}, as understanding the rate of waning can inform the design of appropriate targets and timing for future vaccination programs. Most of the time, waning is assessed by comparing the interval-specific  incidence rate of infection between vaccinated and placebo groups in randomized controlled trials (RCTs). Particularly, an increasing hazard ratio is commonly used as an indication of waning of vaccine efficacy, see for example \cite{durham1998estimation,haber2021comparing,nikas2023estimating}. 
	However, such conventional estimand is misleading because the hazard ratio does not have a causal interpretation due to the inherent selection bias \citep{hernan2010hazards,martinussen2020subtleties,stensrud2019limitations}. As a result, VE may appear to decline over time due to the depletion of individuals most susceptible to infection and would not necessarily mean a waning of immunity. On the other hand,  a comparison of survival curves, i.e. the cumulative vaccine efficacy, does not address waning of immunity and can also be misleading as illustrated in Figure~\ref{fig:surv_waning}. Recently, \cite{janvin2024quantification} introduced  the challenge effect, which is a comparison of incidence rates under a hypothetical intervention on both the vaccination status and the exposure to the  infections agent. They derive bounds on the challenge effect from standard RCT data and propose to use such bounds to quantify waning.  
	
 More in general, the  reduction of the efficacy of a treatment over time is particularly significant in health technology assessments (HTAs), where long-term effectiveness is crucial for evaluating the cost-effectiveness of medical interventions. Evaluation of cost-effectiveness requires extrapolation of the survival functions beyond the study period and such extrapolation can be done in different ways. This generates debates about whether there is treatment effect waning, 
	as neglecting it can lead to overestimation of  cost-effectiveness of treatments. To address this,  guidelines of the National Institute for Health and Care Excellence (NICE) require that companies should provide sensitivity and scenario analyses, including the potential for treatment effect waning, in their economic analysis. There are however no guidelines on how to implement such assumption. Reviews of NICE technology appraisals \citep{trigg2024treatment,armoiry2022assumption,taylor2024treatment,kamgar2022ee228} revealed diverse methods for incorporating treatment effect waning assumptions and also heterogeneity in justifications of considering or disregarding the possibility of waning. 
	
		In this paper, we discuss the possibility to detect and quantify treatment effect waning from time-to-event data. We provide an overview of existing approaches and emphasize once again that the commonly used hazard ratio is not an appropriate metric. 
		In order to assess waning of treatment effect, one would need to choose an estimand which is a causal comparison between treatment and control group at a given time point, and then guarantee that change of such estimand over time reflects waning or strengthening of treatment effect. Hence, {the two groups }
		being compared need to be  similar  both at a fixed time 
{and across} different time points. For comparison of the two groups at fixed time points, we consider recently defined causal versions of the hazard ratio based on principal stratification or the controlled direct effect approach. However, we illustrate that an increase in causal hazard ratios towards one does not necessarily mean {that} the protective effect of the treatment is fading.  Furthermore, the same survival functions may correspond to both scenarios with and without waning, which shows that treatment effect waning cannot be identified from standard time-to-event data without strong untestable
	modeling assumptions. As a consequence, we suggest that the most appropriate way to account for waning in cost-effectiveness analysis is to consider partial identifiability bounds, as a quantification of uncertainty. 
	
	The paper is organized as follows. In Section~\ref{sec:setting} we describe the setting in consideration and introduce the required notation. A more detailed literature review on the topic of waning of vaccine efficacy, including the challenge effect proposed by \cite{janvin2024quantification} is provided in Section~\ref{sec:review}.
	Then, in Section~\ref{sec:cHR}, we discuss two different formulations of causal hazard ratios and show that the challenge effect corresponds to the controlled direct effect in the scenario where, without intervention on exposure, everyone would be exposed to the infectious agent. 
	In Section~\ref{sec:waning} we illustrate through an example that the causal hazard ratios can provide misleading information on waning of treatment effect, which cannot be identified without strong modeling assumptions. Following this, Section~\ref{sec:HTAs} touches upon the issue of extrapolating survival functions for cost-effectiveness analysis under the waning assumption. Finally we conclude with a discussion in Section~\ref{sec:discussion}. 
		
	\section{Notation and setting description}
	\label{sec:setting}
		Throughout this paper we assume to be in the setting of a randomized controlled trial and observe time-to-event data. Particularly, let $X$ denote the time until the event of interest, which could be infection after vaccination, death etc.. Let $A\in\{0,1\}$ denote the randomly assigned treatment group, with $A=0$ corresponding to  the control group.  The observations consist of  $n$ independent identically distributed triplets $(A_i,T_i,\Delta_i)$, where $T$ is the follow-up time, i.e. the minimum between the event time $X$ and the censoring time $C$, and $\Delta=\1{\{X\leq C\}}$ indicates whether the event happened or not. For simplicity we assume independent censoring, which is for example the case when loss of follow-up happens because of end of the study, and consider discrete times, i.e. $X,T\in\{t_1,t_2,\dots\}$. Note that, even though the  event times might be continuous, in practice they are usually observed over a discrete grid. In that case, $X=t_i$ would mean that the event happened during the interval $(t_{i-1},t_{i}]$, with $t_0=0$. For simplicity, we also assume that these time intervals have equal length and therefore, without loss of generality, we can suppose that $X,T\in\{1,2,\dots\}$.
	
\begin{figure}
		\centering
		\includegraphics[width=0.25\textwidth]{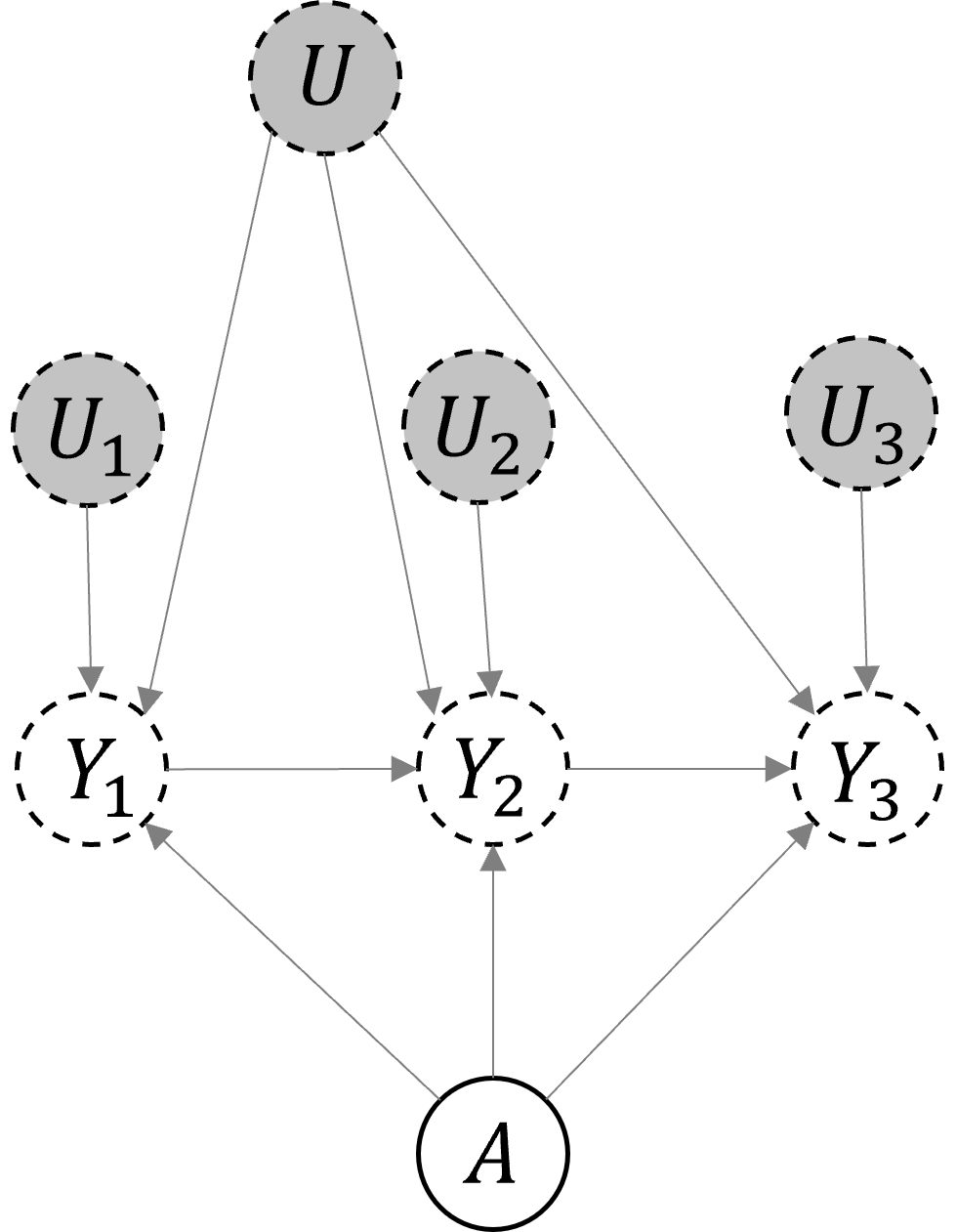}
	\qquad\qquad\qquad \includegraphics[width=0.25\textwidth]{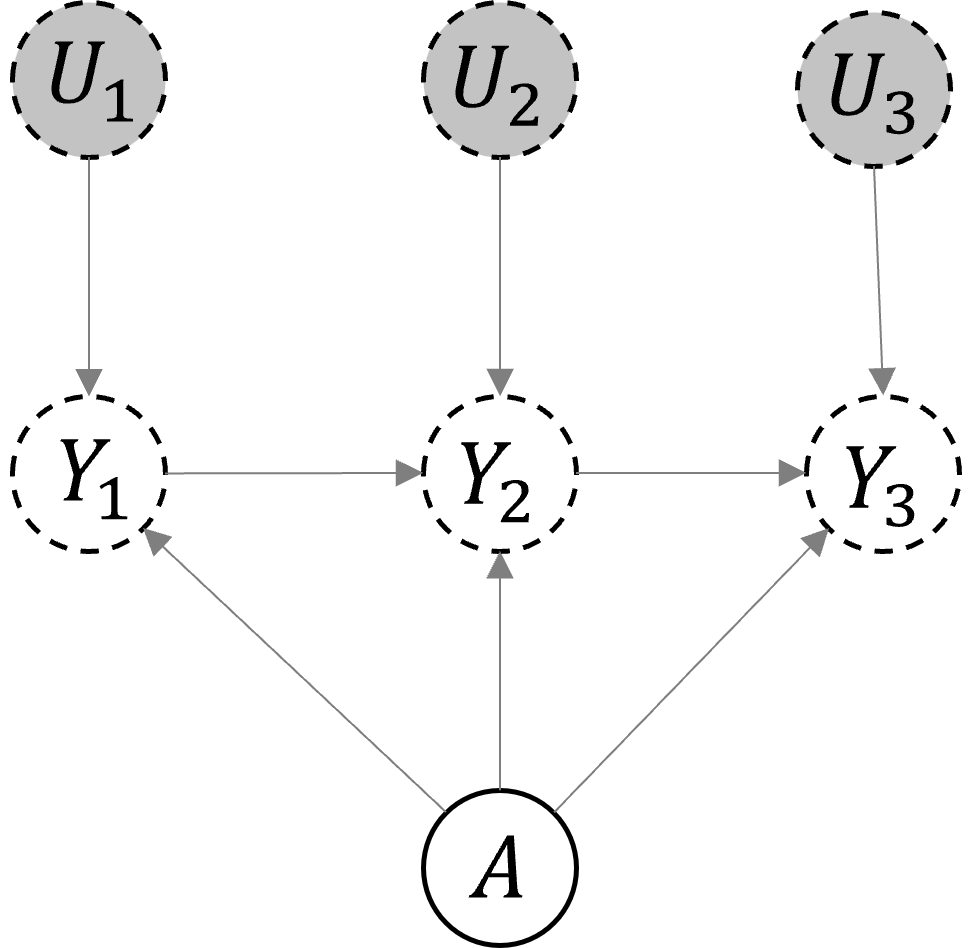}
		\caption{{A general causal DAG (left) and a restricted causal DAG (right) illustrating possible data generating mechanisms. Latent variables are marked in gray {and dashed nodes represent non-intervenable variables.}}}
		\label{fig:general_DAG}
	\end{figure}

	For each time point $t\in\{1,2,\dots\}$, we define the time specific outcome $Y_t=\1{\{X>t\}}$ indicating whether the subject has survived, i.e. has not experienced the event of interest, within  time $t$.  We assume that the event of interest happens only once, or otherwise we are just interested in the first occurrence. Hence, we have  that  $Y_t=0$ implies $Y_s=0$ for all $s>t$  and $Y_t=1$ implies $Y_s=1$ for all $s<t$. 
Let $U$ denote a latent variable which affects survival at different time points or represents latent selection of the study population  \citep{chen2024modeling}. The corresponding causal directed acyclic graph (DAG) illustrating the data generating mechanism is shown in the left panel of Figure~\ref{fig:general_DAG}, where the latent variables $U_i$ are independent random variables affecting only one $Y_i$. The dashed nodes represent non-intervenable variables, as suggested for example in \cite{chen2024modeling}. Note that it is not clear in general how one can directly intervene on $Y_t$, but even if that {were} possible, it would still not be possible to intervene on $Y_s$ for $s>t$ once $Y_t=0$. For our purposes, it will be sufficient considering two time points but the argument
and the DAG can be generalized to more time-points. Note that we are not imposing any specific constraints apart from the natural deterministic relations between the time-specific outcomes and the fact that treatment is randomized. For comparison, we also consider the restricted DAG shown in the right panel of Figure~\ref{fig:general_DAG}, where there is no latent variable or latent selection introducing dependence between the $Y_i$s. Despite not being realistic, this is used as an example of a simplified scenario where the standard hazard ratio has a causal interpretation and it is possible to detect waning of treatment effect.
The notation $Y^a_t$, $X^a$ will be used to denote the potential outcomes for $a=0,1$. Throughout the paper we will focus on the scenario that treatment is protective and waning would refer to a decrease over time in the level of protection provided by the treatment. There is currently no clear, unambiguous definition of waning. The concept will be investigated further in the following sections. 

	The commonly used hazard function for each treatment arm, in the discrete time setting, is 
	 given by
	\begin{equation}
		\label{eqn:hazard}
	\lambda_a(t)={\p(X^a=t\mid X^a\geq t)=\p(Y_t^a=0\mid Y_{t-1}^a=1)}, \text{ a}=0,1.
	\end{equation}
	{Throughout the paper we assume positivity, consistency and treatment randomization, which imply that the hazard function is also given by}
	\begin{equation}
		\label{eqn:hazard2}
		\lambda_a(t)=\p(X=t\mid X\geq t, A=a)=\p(Y_t=0\mid Y_{t-1}=1, A=a), \text{ a}=0,1.
	\end{equation}
	Note that, {under the  independent censoring assumption (i.e. $C\perp X\mid A$), the hazard function can be identified from the observed censored data: }
	\[
	\lambda_a(t)=\p(T=t,\Delta=1\mid T\geq t, A=a).
	\]
However, despite the initial randomization, conditioning on survivors up to a certain time  introduces selection bias and as a result the hazard ratio $\text{HR}(t)=\lambda_1(t)/\lambda_0(t)$ loses the interpretation as a causal contrast for $t>1$. This is because, as it can be seen from the DAG in the left panel of Figure~\ref{fig:general_DAG},  conditioning on $Y_{t-1}$ opens the path between $U$ and $A$. Consequently, changes in $Y_t$ between the two groups can be due to the treatment or to the different distribution of $U$ among the two treatment groups. Such problem has been pointed out in several papers    \citep{hernan2010hazards,martinussen2020subtleties,stensrud2019limitations}. 
	
\begin{figure}
		\includegraphics[width=\textwidth]{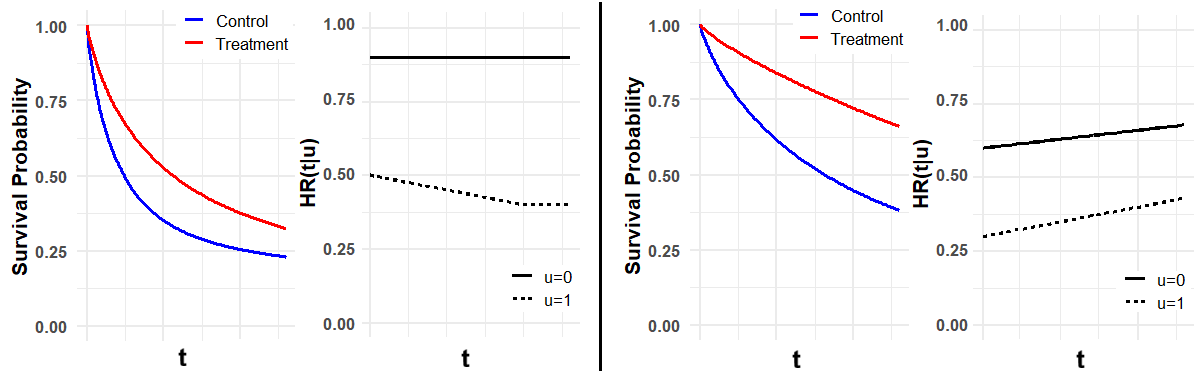}
		\caption{\label{fig:surv_waning}Survival functions for generated RCT data without/with  waning (left/right panel) and the corresponding time varying hazard ratios for the two latent strata of the population ($u=0,1$).} 
	\end{figure}

	Recent literature recommends comparison of survival curves, $$S_a(t)={\p(X^a>t)= }\,\p(X>t\mid A=a),$$ rather than relying on hazard ratios, to draw causal conclusions about treatment effects and inform treatment decisions. However, the survival curve represents a cumulative effect and does not provide a direct understanding of treatment effect waning. Figure~\ref{fig:surv_waning} illustrates that visual inspection of survival curves, even in the continuous time setting, can be misleading regarding the presence of waning. The data for that figure are generated from a population consisting of two latent strata, i.e. $U$ is binary, within which the hazard function is the same. Hence, the hazard ratio conditional on $U$, $\text{HR}(t|u)=\lambda_1(t|u)/\lambda_0(t|u)$, reflects the actual treatment efficacy over time for each of the two types of individuals in the population.  In the left panel treatment effect does not wane for anyone and even gets stronger for some individuals, but from the survival curves one might think that the treatment effect is waning towards the end of the follow-up period {(distance between the curves decreases)}. In the right panel, the efficacy of the treatment decreases for all individuals, but from the survival curves there seems to be no indication of waning  {(distance between the curves is increasing)}. {Note that in this case the problem is not about whether the general or the restricted DAG of Figure~\ref{fig:general_DAG} represents the true data generating mechanism. The same phenomenon happens also in the simple setting of a homogeneous population with just one stratum (restricted DAG): the distance between the two survival functions would always initially increase and then at some point decrease (because they have the same limit 1 at 0 and 0 at infinity) regardless of waning.}
			\section{Literature review on waning of vaccine efficacy}
	\label{sec:review}

	As also discussed in \cite{rubin2022audio}, there is no clear definition of waning of vaccine efficacy, but most think about it as a reduction in a vaccine's ability to provide protection against infection or disease over time and there could be different outcomes of interest such as symptomatic infections, severe infections and death. For example, for COVID 19, vaccines have done better in protecting against severe disease but over time were {believed to be} less protective against infections. 
	Two primary approaches are used to quantify this phenomenon. The first approach uses immunological assays to measure antibody levels post-vaccination, which serve as surrogate markers for vaccine protection. However, such data are rarely available and more expensive to obtain. In addition, another drawback of this method is that it can fail to detect immunity due to resident memory cells, and therefore can be insufficient to fully characterize immunity \citep{rubin2022audio}.  The second and most common approach involves comparing the interval-specific  incidence rate of infection between vaccinated and placebo groups in randomized controlled trials. This method quantifies vaccine efficacy at time $t$ as $$\text{VE}(t)=1-\text{HR}(t)$$ and a decreasing $\text{VE}(t)$ is commonly referred to as waning of vaccine efficacy  \citep{durham1998estimation,haber2021comparing,nikas2023estimating}. In \cite{goldberg2021waning}, it was observed that the rate of  infections
	among people vaccinated at different times increased  as the time
	from vaccination increased in all age groups,
	{both} with and without correction for measured confounding
	factors. This would correspond to an increasing hazard function for the vaccinated individuals. Such  association between the rate of 
	infections and the period of vaccination was used as a  measure of waning immunity with the motivation that, without waning
	of immunity, one would {naively} expect to see no differences
	in infection rates among persons vaccinated
	at different times. 

	However, as argued in Section~\ref{sec:setting}, such conventional VE estimands based on the hazard function or the hazard ratio can be problematic because they compare the hazard rates of populations which differ in the two treatment arms and at different times, lacking therefore a causal interpretation. For example, VE may decline over time due to the depletion of individuals most susceptible to infection and would not necessarily mean a waning of immunity. Addressing this, simple mathematical models, such as the “leaky” (risk of infections is reduced by a constant factor) and “all-or-nothing” (complete or no protection) frameworks, attempt to estimate vaccine efficacy and waning by accounting for unobserved heterogeneity in individual vaccine response \citep{halloran1992interpretation,halloran1997study,smith1984assessment,kanaan2002estimation}. However, these models are overly simplified as they usually assume existence of only a few homogeneous strata among the vaccinated individuals and rely on strong parametric assumptions. Despite this, such studies have illustrated that inferences about the presence of waning cannot be drawn from methods based on the hazard ratio. Particularly, under the leaky model, the estimated
	population hazard ratio in vaccinated individuals relative to
	unvaccinated individuals can appear to increase over time,
	even though vaccine protection does not wane at individual level.
	Furthermore, under the all-or-nothing model, the population
	hazard ratio can appear to remain steady even though individual vaccine protection wanes over time. 
	
\begin{figure}
		\centering
		\includegraphics[width=0.5\linewidth]{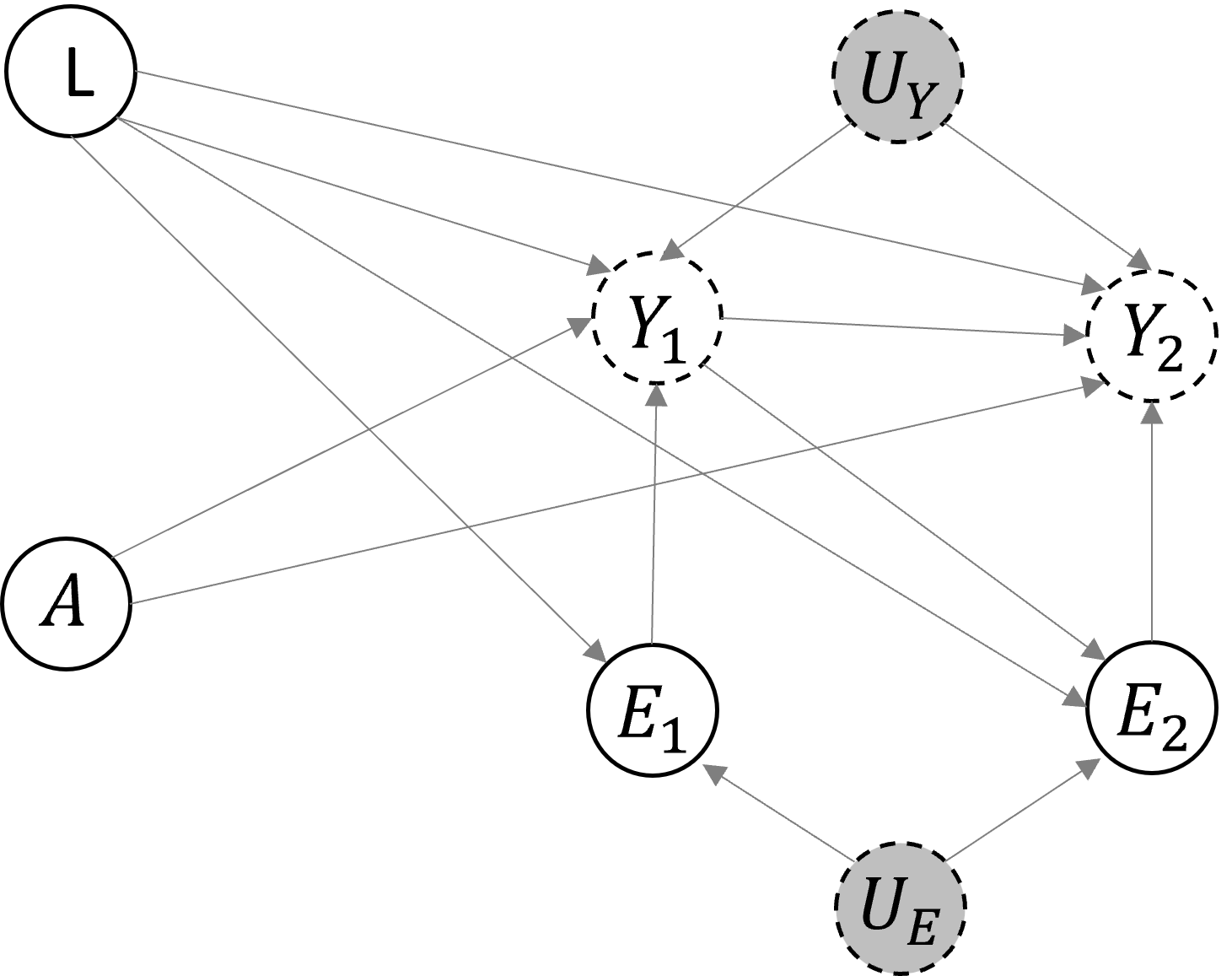}
		\caption{Causal DAG illustrating a data generating mechanism for vaccine effectiveness that satisfies the assumptions of \cite{janvin2024quantification}.}
		\label{fig:DAG_challenge}
	\end{figure}

	To overcome these challenges, a causal “challenge effect” framework has been proposed in \cite{janvin2024quantification}, defining waning in terms of interventions that manipulate both vaccination status $(A)$ and exposure to the infectious agent $(E)$. Conditional on observed baseline covariates $L$, the challenge effect is defined as
	\begin{equation}
		\label{eq:challenge}
		\text{VE}^{ch}(t\mid l)=1-\frac{\p(X^{a=1,\bar{e}_{t-1}=0, e_t=1}=t\mid L=l)}{\p(X^{a=0,\bar{e}_{t-1}=0,e_t=1}=t\mid L=l)},
	\end{equation}
	where $\bar{e}_t=(e_1=\dots=e_{t})$ and  $X^{a,\bar{e}_{t-1}=0, e_t=1}$ denotes the potential outcome under the intervention that assigns treatment $a$, isolates the individual until time $t-1$ and exposes  the individual to the {infectious agent} (in the same controlled way) at time $t$. Then they define vaccine effect waning as a decrease in the challenge effect. Note that the challenge effect coincides with the conventional vaccine efficacy at the first time point $\text{VE}^{ch}(1\,|\,l)=\text{VE}(1\,|\, l)$, but not at future times, unless there are no unobserved variables that affect the survival status at different time points. Although ideal challenge trials, where exposure to the infectious agent is controlled, could identify the challenge effect, such trials are often unethical or impractical. However, \cite{janvin2024quantification} provide bounds for the challenge effect using data from conventional RCTs, where exposure occurs naturally through community interactions and exposure status is not measured. The required assumptions are: consistency; exposure necessity, i.e. infection cannot happen without exposure; no treatment effect on exposure in the unexposed ($E_t^{a=0,\bar{e}_{t-1}=0}=E_t^{a=1,\bar{e}_{t-1}=0})$; exposure exchangeability, i.e. exposure and outcome are unconfounded conditional on measured baseline covariates $L$; exposure effect restriction, i.e. 
	\[
	\begin{aligned}
		\p(X=t\mid A=a,L=l)&\leq \p(X^{\bar{e}_{t-1}=0}=t\mid A=a,L=l)\\
		&\leq \p(X\leq t\mid A=a,L=l).
	\end{aligned}
	\]
 The latter assumption could be violated if there exist causal paths from $E_t$ to $Y_{s}$ for $s>t$ that are not intersected by $Y_t$, for example if previous isolation affects future immunity. As noted by the authors, a DAG for the data generating mechanism that satisfies such assumptions is given in Figure~\ref{fig:DAG_challenge}, where $U_Y$ and $U_E$ denote unobserved variables. Based on the derived partial identification bounds, \cite{janvin2024quantification} propose that one could conclude waning of treatment efficacy if for example the upper bound for $\text{VE}(t+1\,|\,l)$ is lower than the lower bound for  $\text{VE}(t\,|\,l)$.

	\section{Causal hazard ratios }
	\label{sec:cHR}
In order to address the lack of causal interpretation of the hazard ratio, different versions of causal hazard ratios have been proposed in the literature, which allow for a causal comparison of treatment and control group at different time points. Despite not being identifiable from the observed data in absence of strong untestable modeling  assumptions, one could derive nonparametric bounds of such causal hazard ratios and wonder whether they can be used to draw  conclusions about waning of treatment effect.  We describe these approaches below and derive their relation with the challenge effect of \cite{janvin2024quantification}. Since exposure is particular for vaccine effectiveness, which is a more complicated scenario due to the dependence on the prevalence of the virus in the population and changes in the virus, we focus instead on the problem of treatment effect waning in settings where individuals are continuously at risk for the event of interest. This is the same as assuming exposure is a deterministic variable always equal to 1.

\subsubsection*{Causal hazard ratio: principal stratification approach }
		This version of a causal hazard ratio was proposed by \cite{martinussen2020subtleties} and is based on the principal stratification approach \citep{frangakis2002principal}. The discrete version is defined as
	\begin{equation}
		\label{eqn:cHR_1}
		\text{cHR}_{1}(t)=\frac{\p\left(X^1=t \mid X^0\geq t, X^1\geq t\right)}{\p\left(X^0=t\mid X^0\geq t, X^1\geq t\right)},
	\end{equation}
	and compares the risk of the event happening at time $t$ for the treatment and control groups, among those who would survive until time $t$ regardless of the assigned treatment. The principal stratum `LL' defined as $\{X^0\geq t, X^1\geq t\}$ is an unknown subgroup of the population but we can bound the probability of being in this strata. Particularly, if such probability is known to be high, the analysis on the `LL' stratum is relevant because that is the dominant subgroup of the population. 
	
\subsubsection*{Causal hazard ratio: controlled direct effect approach }
	This causal hazard ratio is defined as 
	\begin{equation}
		\label{eqn:cHR_2}
		\begin{aligned}
			\text{cHR}_{2}(t)=\frac{\E[\lambda_1(t\mid U)]}{\E[\lambda_0(t\mid U)]},
		\end{aligned}
	\end{equation}
	{where $U$ represents the latent random variable in the left DAG of Figure~\ref{fig:general_DAG} that affects the survival outcome at different times.}
	{We call this a controlled direct effect approach because, if we would be able to intervene on the survival status by preventing the event of interest from happening until a certain time \citep{aalen2015does} without modifying the distribution of $U$, we would get }
	\begin{equation}
		\label{eqn:cHR_2_inter}
		\begin{aligned}
			\text{cHR}_{2}(t)
			&={\frac{\int\p\left(X^1=t \mid X^1\geq t,U=u\right)\dd{F_U(u)}}{\int\p\left(X^0=t \mid X^0\geq t,U=u\right)\dd{F_U(u)}}}\\
			&={\frac{\p\left(Y_t^{a=1,y_{t-1}=1}=0 \right)}{\p\left(Y_t^{a=0,y_{t-1}=1}=0\right)}.}
		\end{aligned}
	\end{equation}
	However, such abstract intervention is in general not possible in practice and, without clearly defining how the intervention {``$Y_{t-1}=1$"} 
	 is performed, its meaning remains ambiguous. In the particular context of vaccine efficacy, one possible such intervention could be through controlled exposure as in \cite{janvin2024quantification}: by keeping individuals isolated, one prevents infection from happening. 
	We note that $\text{cHR}_2$ is also the same as the discrete version of the causal hazard ratio proposed by \cite{post2024built} in their Definition~1, where a specific structural causal model is considered. It can be interpreted as the ratio of
	the expected hazard rates in the treatment and control group with respect to the initial distribution of the latent variable $U$.
	
	\begin{remark}
		Note that, at the first time point, both these causal hazard ratios coincide with the conventional  hazard ratio $\text{cHR}_1(1)=\text{cHR}_2(1)=\text{HR}(1)$ as a result of the initial randomization. However, they are generally different at subsequent time points.  The conventional  HR can also be written as
		\begin{equation}
			\label{eq:HR}
			\text{HR}(t)=\frac{\p(X^1=t\mid X^1\geq t)}{\p(X^0=t\mid X^0\geq t)}=\frac{\E_{U|{X^1\geq t}}[\lambda_1(t\mid U)]}{\E_{U|{X^0\geq t}}[\lambda_0(t\mid U)]},
		\end{equation}
		where the expectations are taken with respect to the distribution of $U$ among the survivors of {the respective} treatment, which are {in general} different from each other and from the initial distribution of $U$, used for the expectations in \eqref{eqn:cHR_2},  {when treatment has an effect on survival}.  
		On the other hand, the causal hazard ratio based on principal stratification can be written as
		\begin{equation}
			\label{eq:cHR1_2}
			\text{cHR}_1(t)=\frac{\E_{U|X^0\geq t,X^1\geq t}[\lambda_1(t\mid U)]}{\E_{U|X^0\geq t,X^1\geq t}[\lambda_0(t\mid U)]},
		\end{equation}
		where the expectations are taken with respect to the distribution of $U$ among the individuals that would survive until time $t$ under both treatments.
	\end{remark}
\begin{remark}
\label{cHR-restricted_dAG}
If one considers the restricted DAG in the right panel of Figure~\ref{fig:general_DAG}, which can be thought as having just one stratum in the population with hazard $\lambda_a(t)$, the  two causal hazard ratios coincide with the conventional HR. {Indeed,}  $\text{HR}(t)=\text{cHR}_2(t)$ by definition since in this case there is no condition on $U$. Moreover, the corresponding structural causal model is given by 
	\[
	Y_t=f_t(A,Y_{t-1},U_t),\quad t\in{1,2,\dots},
	\] 
	where $A$ and 
	$(U_t)_t$ are exogenous variables all independent of each other, $Y_0=1$ and   $f_t(A,0,U_t)=0$. Hence, since $Y_{t-1}^0,Y_{t-1}^1$ depend only on $(U_s)_{s<t}$ and are independent of $U_t$, we have
	\[
	\begin{aligned}
			\p(Y_t^1=0\mid Y_{t-1}^1=1)&=\p(f_t(1,1,U_t)=0\mid Y_{t-1}^1=1)\\
			&=\p(f_t(1,1,U_t)=0\mid Y_{t-1}^0=1,Y_{t-1}^1=1)\\
			&=\p(Y_t^1=0\mid Y_{t-1}^0=1,Y_{t-1}^1=1)
	\end{aligned}
	\] 
	It follows that
\[
\text{HR}(t)=\frac{\p(Y_t^1=0\mid Y_{t-1}^1=1)}{\p(Y_t^0=0\mid Y_{t-1}^0=1)}=\frac{\p(Y_t^1=0\mid Y_{t-1}^0=1,Y_{t-1}^1=1)}{\p(Y_t^0=0\mid Y_{t-1}^0=1,Y_{t-1}^1=1)}=\text{cHR}_1(t).
\]
In particular this means that the conventional HR has a causal interpretation as the principal stratum causal effect or the controlled direct effect. Additionally, in such case, an increasing HR would have a waning interpretation. We can assume without loss of generality that $U_t\in\{$`00',`01',`10',`11'$\}$, where $U_t=$`$cd$' means that $Y_t^0=c$ and $Y_t^1=d$ if $Y_{t-1}^1=Y_{t-1}^0=1$. Then we obtain
\[
\text{HR}(t)=\frac{\p(U_t=\text{`$00$' or }U_t=\text{`$10$'})}{\p(U_t=\text{`$00$' or }U_t=\text{`$01$'})}.
\]
The numerator and denominator represent the proportion of individuals experiencing the event at time $t$ when treated or non-treated respectively, if they had all survived until that point. 
An increasing HR would mean that the protection of the treatment (in terms of reducing the proportion of individuals experiencing the event) decreases over time.
	\end{remark}
	Below we show that, under the `always exposed scenario', 
	the challenge effect of \cite{janvin2024quantification} corresponds to $\text{cHR}_2$.
Consider the DAG in Figure~\ref{fig:DAG_challenge}, which satisfies the assumptions of \cite{janvin2024quantification}, assuming that $E_t=1$ with probability one. We ignore for now the observed covariates $L$, as the argument would not change even conditioning on them. A structural causal model compatible with this DAG is given by
\[
\begin{aligned}
Y_t&=f_t(A,Y_{t-1},U_Y,E_t,U_t),
\\
E_t&=g_t(U_E,V_t),  \qquad\qquad t\in\{1,2,\dots\}
\end{aligned}
\]
where $A, (U_t)_t, U_Y, {U_E, (V_t)_t}$ are independent exogenous variables, {$g_t$ is a constant function equal to 1,} 
 $Y_0=1$ and the functions $f_t$ satisfy  the following relations $$f_t(A,0,U_Y,E_t,U_t)=0,\qquad f_t(A,1,U_Y,0,U_t)=1.$$  
 Since intervention on $E_t$ and $A$ does not affect the distribution of $U_Y$, we have that
		\[
		\begin{aligned}
			&\p(X^{a,\bar{e}_{t-1}=0,e_t=1}=t)={\p(X^{a,\bar{e}_{t-1}=0}=t)}\\
			&=\int {\p\left(X^{a,\bar{e}_{t-1}=0}=t\mid U_Y=u\right)}\dd F_{U_Y}(u)\\
&=\int {\p\left(Y_t^{a,\bar{e}_{t-1}=0}=0, Y_s^{a,\bar{e}_{t-1}=0}=1 \text{ for all }s<t\mid U_Y=u\right)}\dd F_{U_Y}(u)\\
			&=\int {\p\left(Y_t^{a,\bar{e}_{t-1}=0}=0 \mid U_Y=u,Y_s^{a,\bar{e}_{t-1}=0}=1 \text{ for all }s<t\right)}\\	
			&\qquad \times{\p\left(Y_s^{a,\bar{e}_{t-1}=0}=1 \text{ for all }s<t\mid U_Y=u\right)}\dd F_{U_Y}(u).
				\end{aligned}
			\]
		The second probability in the integrand is 1 due to the exposure necessity assumption. {In addition, 
		for the assumed SCM	we have
			\[
			\begin{aligned}
&\p\left(Y_t^{a,\bar{e}_{t-1}=0}=0 \mid U_Y=u,Y_s^{a,\bar{e}_{t-1}=0}=1 \text{ for all }s<t\right)\\
&=\p\left(f_t(a,Y_{t-1}^{a,\bar{e}_{t-1}=0},U_Y,1,U_t)=0\mid U_Y=u,Y_s^{a,\bar{e}_{t-1}=0}=1 \text{ for all }s<t\right)\\
&=\p\left(f_t(a,1,u,1,U_t)=0\right)\\
&=\p\left(f_t(a,1,u,1,U_t)=0\mid f_s(a,1,u,1,U_s)=1 \text{ for all }s<t\right)\\
&=\p\left(f_t(a,Y_{t-1}^{a},U_Y,1,U_t)=0\mid U_Y=u,Y_s^{a}=1 \text{ for all }s<t\right)\\
&=\p\left(Y_t^{a}=0 \mid U_Y=u,Y_s^{a}=1 \text{ for all }s<t\right).
			\end{aligned}
			\]
			It follows that}
		\[
		\begin{aligned}
			&\p(X^{a,\bar{e}_{t-1}=0,e_t=1}=t)\\
			&=\int {\p\left(Y_t^{a,\bar{e}_{t-1}=0}=0 \mid Y_s^{a,\bar{e}_{t-1}=0}=1 \text{ for all }s<t, U_Y=u\right)}\dd F_{U_Y}(u)\\
			&=\int \p\left(Y^{{a}}_t=0 \mid Y^{{a}}_s=1 \text{ for all }s<t,U_Y=u\right)\dd F_{U_Y}(u)\\
			&=\E[\lambda_a(t|U_Y)].
		\end{aligned}
		\]
	Hence, in this context, the challenge effect defined in~\eqref{eq:challenge} can be written in terms of the causal hazard ratio based on the controlled direct effect approach defined in \eqref{eqn:cHR_2}:
	\[
	\text{VE}^{ch}(t)=1-\text{cHR}_2(t).
	\]
	
	\section{Discussion on identification of  treatment effect waning}
	\label{sec:waning}

	In this section  we illustrate that both the   causal hazard ratios defined in Section~\ref{sec:cHR}, and consequently also the challenge effect proposed by \cite{janvin2024quantification} {when assuming the data generating process of Figure~\ref{fig:DAG_challenge} and the `always exposed scenario'}, are not appropriate for characterizing the waning of the treatment effect. Moreover, the purpose is to show that the observed data alone do not provide information on treatment effect waning without strong modeling assumptions.   Particularly, we consider the case in which $\text{HR}(1)<1$, 
	meaning that initially the treatment has a protective effect, but an increasing causal hazard ratio does not mean that the treatment effect is waning and vice versa. 
	{In addition, the following example illustrates that {there exist two} 
		survival functions $S_1$ and $S_0$  for which two opposite scenarios are possible: the treatment effect at individual level  wanes  for everyone, or it becomes stronger for everyone {over time} (in particular, it does not wane). This means  in particular  that, whatever estimand we use to characterize waning, we  would not be able to tell whether
		treatment effect is waning at individual level (on average) by looking at the observed survival functions. }
	
\begin{figure}
\includegraphics[width=\textwidth]{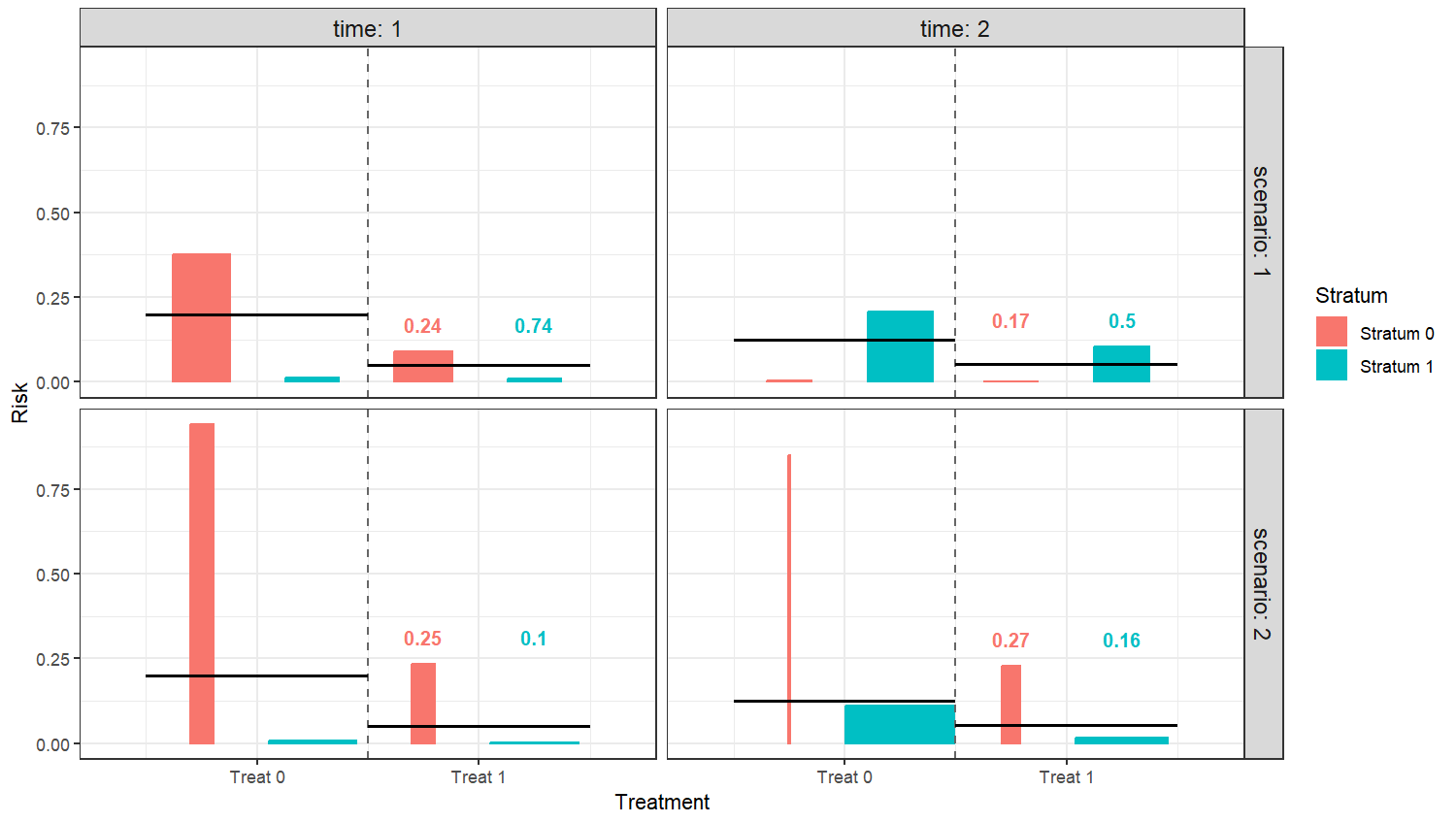}
\caption{\label{fig:scenarios} {Visualization of the two scenarios in Example~\ref{example:scenarios}: risk by treatment group, time and stratum. The width of the bars corresponds to the stratum proportion. The horizontal black line corresponds to the overall (combined) risk. The numbers on top of the bars for treatment 1 represent the amount of risk reduction that treatment offers with respect to the control group for that specific stratum, i.e. HR$(t|u)$.}}
	\end{figure}

	\begin{example} 
		\label{example:scenarios}
		Consider the following simple data generating mechanism, which follows the left DAG in Figure~\ref{fig:general_DAG}, or the one in Figure~\ref{fig:DAG_challenge} under the `always exposed' scenario and no observed baseline covariates. Treatment $A$ is randomly generated as a Bernoulli variable with probability $0.5$ as in a randomized trial. The population consists of two hidden strata defined by a latent baseline variable $U\sim\text{Bernoulli}(\rho)$ (same as $U_Y$ in Figure~\ref{fig:DAG_challenge}), which determines the hazard rate at all time points $\lambda_a(t|u)$. Then, for each stratum, if $Y^a_{t}=1$, the survival status at the next time point is generated as a Bernoulli variable with probability $\lambda_a(t{+1}|u)$, independently for the two potential outcomes. {Specifically, $Y^a_{t+1}=\1_{(\lambda_a(t+1|u),1]}(U^a_{u,t+1})$, where $(U_{0,1}^0,U_{0,1}^1,U_{0,2}^0,U_{0,2}^1,\dots)$, $(U_{1,1}^0,U_{1,1}^1,U_{1,2}^0,$ $U_{1,2}^1,\dots)$ are all i.i.d. variables with uniform distribution in $(0,1)$. With respect to the notation in Figure~\ref{fig:general_DAG}, $U_t=(U^0_{0,t},U^1_{0,t},U^0_{1,t},U^1_{1,t}).$} Note that this is always a possible underlying data generating mechanism for the observed data. In such case, the unobserved $\text{HR}(t|u)$ represents the individual treatment effect and an increasing $\text{HR}(t|u)$ over time would mean that treatment effect is waning for individuals corresponding to $U=u$. \\
		{
			Consider a setting in which the survival functions, at two time points, are given by:			
		\begin{equation}
			\label{eqn:obs_surv}
			\begin{cases}
				S_0(1)&=0.8\\
				S_1(1)&=0.95
			\end{cases},\qquad
			\begin{cases}
				S_0(2)&=0.7\\
				S_1(2)&=0.9.
			\end{cases}
			\end{equation}
			Note that $S_1-S_0$ increases over time, which might at first sight suggest that treatment effect does not wane. Using~\eqref{eqn:hazard} we can compute the resulting hazard functions and hazard ratios: 
			\[
			\begin{cases}
				\lambda_0(1)&=0.2\\
				\lambda_1(1)&=0.05
			\end{cases},\qquad
			\begin{cases}
				\lambda_0(2)&=0.125\\
				\lambda_1(2)&=0.053
			\end{cases}
			,\qquad
			\begin{cases}
				\text{HR}(1)&=0.25\\
				\text{HR}(2)&=0.42
			\end{cases}
			\]
			The HR would indicate that there is waning of treatment effect since $\text{HR}(1)<\text{HR}(2)$. However, at individual level, the following two scenarios are possible, i.e. compatible with the observed survival functions.}\\
		
		{
			\textbf{Scenario 1 (no waning for all)}\\
			We take $\rho=0.48$. The individual hazard rates under the control group are given by: 
			\[
			\begin{cases}
				\lambda_0(1\mid U=0)&=0.377\\
				\lambda_0(1\mid U=1)&=0.012
			\end{cases},\qquad
			\begin{cases}
				\lambda_0(2\mid U=0)&=0.005\\
				\lambda_0(2\mid U=1)&=0.206
			\end{cases}
			\]
			and the individual hazard ratios are given by:
			\[
			\begin{cases}
			\text{HR}(1\mid U=0)&=0.24\\
			\text{HR}(1\mid U=1)&=0.74
			\end{cases},\qquad
			\begin{cases}
			\text{HR}(2\mid U=0)&=0.17\\
			\text{HR}(2\mid U=1)&=0.50.
			\end{cases}
			\]
			To check that such scenario is compatible with the observed survival functions one needs to check the following relations
			\begin{equation}
				\label{eqn:surv-hazard}
				\begin{aligned}
				\rho\lambda_0(1|1)+(1-\rho)\lambda_0(1|0)&=1-S_0(1)\\
				\rho\text{HR}(1|1)\lambda_0(1|1)+(1-\rho)\text{HR}(1|0)\lambda_0(1|0)&=1-S_1(1)\\
				\rho_2^0\lambda_0(2|1)+(1-\rho_2^0)\lambda_0(2|0)&=\frac{S_0(1)-S_0(2)}{S_0(1)}\\
				\rho^1_2\text{HR}(2|1)\lambda_0(2|1)+(1-\rho^1_2)\text{HR}(2|0)\lambda_0(2|0)&=\frac{S_1(1)-S_1(2)}{S_1(1)},\\
				\end{aligned}
				\end{equation}
				where 
				\[
				\rho_2^k=\p(U=1\mid X^k\geq 2)=\frac{\rho[1-\text{HR}(1|1)^k\lambda_0(1|1)]}{S_k(1)}, \quad k=0,1.
				\]
			From Remark 1 we have that $\text{cHR}_1(1)=\text{cHR}_2(1)=\text{HR}(1)=0.25$. Moreover, using~\eqref{eqn:cHR_2}, we can compute
			\begin{equation}
\label{eqn:comp_cHR2}
			\begin{aligned}
\text{cHR}_2(2)&=\frac{(1-\rho)\text{HR}(2|0)\lambda_0(2|0)+\rho\text{HR}(2|1)\lambda_0(2|1)}{(1-\rho)\lambda_0(2|0)+\rho\lambda_0(2|1)}=0.49.
			\end{aligned}
		\end{equation}
		For $\text{cHR}_1(2)$ we first compute the distribution of $U$ among the `always survivors' principal strata `LL':
		\begin{equation}
			\label{eqn:comp_U}
			\begin{aligned}
				\rho_2^{LL}&:=\p(U=1\mid X^0\geq 2, X^1\geq 2)\\
				&=\frac{\p(X^0\geq 2, X^1\geq t\mid U=1)\p(U=1)}{\p(X^0\geq 2, X^1\geq 2)}\\
				&=\frac{\p(X^0\geq t\mid U=1)\p(X^1\geq2\mid U=1)\p(U=1)}{\sum_{k=0}^1\p(X^0\geq 2|U=k)\p( X^1\geq 2\mid U=k)\p(U=k)}\\
				&=\frac{\rho \{1-\text{HR}(1|1)\lambda_0(1|1)\}\{1-\lambda_0(1|1)\}}{\sum_{k=0}^1\rho^k(1-\rho)^{1-k} \{1-\text{HR}(1|k)\lambda_0(1|k)\}\{1-\lambda_0(1|k)\}}=0.62.
			\end{aligned}
		\end{equation}
		As a result, from \eqref{eq:cHR1_2}, we obtain
		\begin{equation}
			\label{eqn:comp_cHR1}
			\begin{aligned}
				\text{cHR}_1(2)&=\frac{(1-\rho_2^{LL})\text{HR}(2|0)\lambda_0(2|0)+\rho_2^{LL}\text{HR}(2|1)\lambda_0(2|1)}{(1-\rho_2^{LL})\lambda_0(2|0)+\rho_2^{LL}\lambda_0(2|1)}=0.50.
			\end{aligned}
		\end{equation} 
			In this scenario, based on the causal hazard ratios it also seems  like treatment effect wanes ($\text{cHR}_1(1)<\text{cHR}_1(2), \text{cHR}_2(1)<\text{cHR}_2(2)$), but actually it does not wane for anyone ($\text{HR}(1|U=k)>\text{HR}(2|U=k)$ for both $k=0,1$){; rather, the treatment becomes more effective over time for everyone.}}\\
		
		{
			\textbf{Scenario 2 (waning for all)} \\
			We take $\rho=0.79$. The individual hazard rates under the control group are: 
			\[
			\begin{cases}
				\lambda_0(1\mid U=0)&=0.94\\
				\lambda_0(1\mid U=1)&=0.009
			\end{cases},\qquad
			\begin{cases}
				\lambda_0(2\mid U=0)&=0.85\\
				\lambda_0(2\mid U=1)&=0.11,
			\end{cases}
			\]
			and the individual hazard ratios are:
			\[
			\begin{cases}
			\text{HR}(1\mid U=0)&=0.25\\
				\text{HR}(1\mid U=1)&=0.10
			\end{cases},\qquad
			\begin{cases}
			\text{HR}(2\mid U=0)&=0.27\\
			\text{HR}(2\mid U=1)&=0.16.
			\end{cases}
			\]
				One can check that such scenario is compatible with the observed survival functions in~\eqref{eqn:obs_surv} using~\eqref{eqn:surv-hazard}.
			Then we have that $\text{cHR}_1(1)=\text{cHR}_2(1)=\text{HR}(1)=0.25$ and, as in \eqref{eqn:comp_cHR2}-\eqref{eqn:comp_cHR1}, we can compute $\text{cHR}_1(2)=0.17$, $\text{cHR}_2(2)=0.23$. 
			In this scenario, based on the causal hazard ratios it seems like treatment effect doesn't wane ($\text{cHR}_1(1)>\text{cHR}_1(2), \text{cHR}_2(1)>\text{cHR}_2(2)$), but it actually wanes for all individuals ($\text{HR}(1|U=k)<\text{HR}(2|U=k)$ for both $k=0,1$). }
	\end{example}
	We visualize both scenarios in Figure~\ref{fig:scenarios}. It is clear that the initial distribution of the two strata among the two treatment groups is the same (mimicking a randomized study). For each treatment group and each time point, the risk of the two strata is different among the two scenarios. However, the overall risk, which is a weighted average depending on the relative size of each stratum, {is} the same for the two scenarios, making them indistinguishable based on only the observed data. For example, for $A=0$ and $t=1$, in the first scenario the size of the two strata is almost the same, while in the second scenario the higher risk stratum is smaller but at the same time its risk is higher. Overall, the percentage of the subjects who experience the event ends up being the same for both scenarios. At time $t=2$, the protection offered by the treatment strengthens in scenario 1 and weakens in scenario 2 for both strata, but given that they start with different risks under $A=0$, it is still possible that the overall risk under treatment at time $t=2$ is the same under both scenarios.	{At first sight, it might seem unrealistic that in scenario 1 the risk for stratum 0 drops to nearly 0 at time 2, while the risk for stratum 1 rises from nearly  0 to 0.2. However, distinct sub-populations can in general exhibit very different risk trajectories. For instance, a virus may evolve over time, and different variants can disproportionately affect different groups. }	  
	
	 {The phenomenon illustrated in Example~\ref{example:scenarios} might resemble Simpson's paradox, or more generally the reversal paradox \citep{tu2008simpson}. This paradox occurs when the association between two variables changes direction  after conditioning on a third variable. For example, a treatment can appear effective within each subgroup (e.g., men and women separately), but ineffective or even harmful when data are aggregated. Then, the main question that arises is whether one should use the conditional or unconditional analysis to draw causal conclusions from the data. Several papers have discussed settings under which the Simpson's paradox can arise. Most often the underlying cause of the paradox is confounding, i.e. the conditioning variable is distributed differently across treatment groups, but the paradox could occur for example also when conditioning on a mediator or when using non collapsible association measures \citep{hernan2011simpson,Pearl02012014,dong2024simpson}. In our setting, the problem is more complex since we are not just concerned about the association between two variables, but whether the strength of the direct association of treatment and $Y_t$ decreases over time. The key message is also different: two opposite scenarios of treatment effect waning are indistinguishable from the observed data because the stratifying variable $U$ is unobserved and cannot be adjusted for. In particular, whatever definition of waning we adopt from the observed data,  conditioning on $U$ (if it was observed) could reverse conclusions about waning.  While $U$ is not a standard confounder since it does not affect treatment assignment, it still confounds the direct effect of treatment on $Y_t$ when conditioning on $Y_{t-1}$.}
	 
		Note that Example~\ref{example:scenarios} corresponds to a simplified data generation mechanism and population structure. However, it is sufficient for illustrating the message that, since both scenarios are probable, the causal hazard ratios and any other estimand we could target, would lead to the wrong conclusion in one of the scenarios. We on purpose chose to construct the two underlying scenarios in the extreme cases that there is waning for everyone or for no one, because then it is clear and unambiguous that the resulting conclusions are wrong. Otherwise one would need to argue in terms of expectations which could be defined in different ways.  In addition,  we are not concerned with the partial identifiability of the causal hazard ratios and the additional estimation uncertainty, as a result of having only a finite sample and not knowing the true survival functions, because the issue arises already in the ideal world where all the true quantities are known. 
		
		Intuitively, the causal hazard ratio defined in terms of the principal strata is not appropriate for understanding waning of the treatment effect because the composition of the principal strata changes over time (distribution of $U$), hence the populations at time $t$ and $t+1$ are different and not comparable with each other. On the other hand, the causal hazard ratio based on the controlled direct effect, is the ratio of the expected hazard rates with respect to the initial distribution of $U$ representing the composition of the population at time zero. So, the population we are looking at remains the same at all times, but it still doesn't properly reflect waning because it is a ratio of expected hazards instead of the expectation of the ratio. In addition, the expectation with respect to the initial distribution of $U$ might not be the group of interest at future times, when specific subgroups might be the majority of the ones who survive.  {To solve these problems,} one could think of more appropriate target estimands such as 
	\[
	\E[\text{HR}(t\mid U)]\qquad\text{ or }\qquad\E[\text{HR}(t\mid U)\mid Y_{t}^1=1],
	\]
	which can also only be partially identified. 
	However, in situations where the same survival functions correspond to both settings with and without waning {(actually even strengthening of treatment effect)} for all individuals as seen in Example~\ref{example:scenarios}, the identification bounds of such estimands would not be informative in terms of treatment effect waning. This is because the observed data cannot exclude any of the possible scenarios. We suspect that this situation is likely to occur most frequently in practice.  
	
	Conditioning on observed covariates would not solve the problem {either}  because it is {typically} impossible to observe all variables that affect survival at different time points. Note that the situation is different from average treatment effect estimation with  observational data settings where  one outcome is observed and one hopes that, by conditioning on baseline covariates, the treatment assignment becomes independent of the outcome. {In such case, one just needs to identify covariates that affect treatment decision so knowledge about how such decision is made would solve the problem. Instead, here we need all possible covariates that affect survival over time and it is very unlikely to possess such knowledge in practice.} Assuming strict monotonicity in terms of potential outcomes, i.e. $X^1\geq X^0$, meaning that treatment can only prolong the survival time almost surely, would also not help since the dependence between the potential outcomes only matters for $\text{cHR}_1$, which as explained above cannot be used to infer waning, because the reference population changes among different times. If one would assume that for all individuals the baseline hazard $\lambda_0(t|u)=c(u)$ is constant over time, an increasing $\text{cHR}_2$ would mean that $\E[\text{HR}(t|U)\lambda_0(t|U)]$ is increasing {(see \eqref{eqn:cHR_2})}. Then we could deduce that at least for some individuals in the population, there is treatment effect waning. However, it could still be that there is waning only for a minority of the population for which $c(u)$ is larger. {If instead we would make a stronger assumption that treatment effect is homogeneous across all individuals, i.e. $\text{HR}(t|u)$ is constant with respect to $u$, then from \eqref{eqn:cHR_2}, \eqref{eq:HR} and \eqref{eq:cHR1_2}, it follows that the three notions of hazard ratios coincide with the individual hazard ratio $\text{cHR}_2(t)=\text{cHR}_1(t)=\text{HR}(t)=\text{HR}(t|u)$. In particular, an increasing HR could then be interpreted as a waning of treatment effect. The assumption of homogeneous treatment effect is however untestable.}

	\section{Cost effectiveness analysis under the waning assumption}
	\label{sec:HTAs}
	
\begin{figure}
		\centering
		\includegraphics[width=0.5\linewidth]{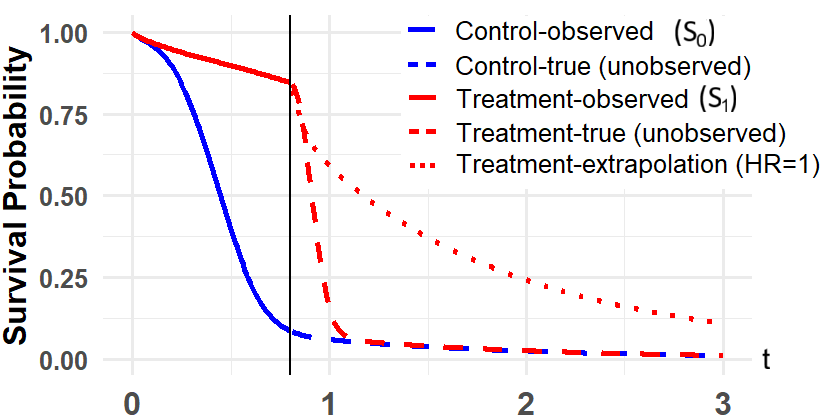}
		\includegraphics[width=0.46\linewidth]{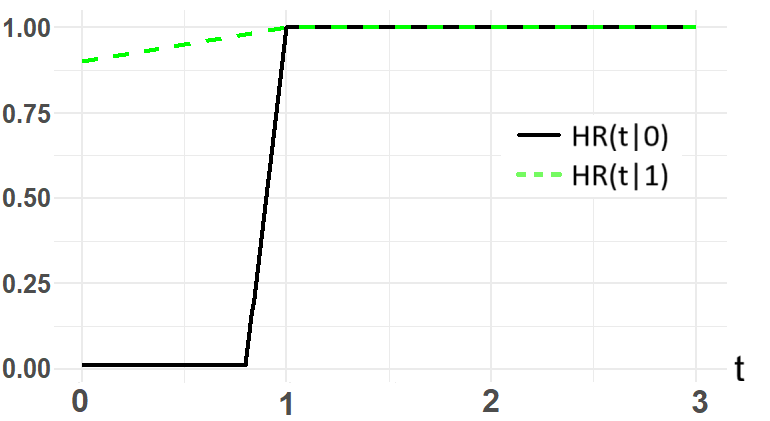}
		\caption{Left: Survival functions for the scenario described in Example~\ref{example:surv}. Study duration is $\tau=0.8$ and the dotted red line corresponds to the extrapolated survival function for the treatment group under the assumption that $\text{HR}(t)=1$ for $t>\tau$. {Right: True underlying hazard ratio for the two strata of the population.}} 
		\label{fig:surv_extrapolation}
	\end{figure}

	In the context of  cost-effectiveness  for  health technology assessments (HTAs), the interest is not only in understanding whether there is treatment effect waning, but also on extrapolating the survival function beyond the study duration under {a} waning assumption for worst-case scenario analysis. 
	 There are however no guidelines on how to implement such assumption. Reviews of NICE technology appraisals \cite{trigg2024treatment,armoiry2022assumption,taylor2024treatment,kamgar2022ee228} revealed diverse methods for incorporating treatment effect waning assumptions and also heterogeneity in justifications for {considering or disregarding} the possibility of waning. Sometimes 
	 analysis of treatment discontinuation is considered 
	 as a proxy for waning of the treatment effect with the motivation that if patients notice loss of efficacy they would discontinue the treatment.  However,  treatment discontinuation could happen also for other reasons not related to effect waning, such as deterioration of the patient's condition. 
{Under the assumption that treatment is not detrimental,} incorporation of treatment effect waning into the model is done by imposing that either the hazard (the most common) or the survival functions of the two groups become equal after a period of time. Such adjustment could be applied gradually or suddenly. Example~\ref{example:surv} below and  Figure~\ref{fig:surv_extrapolation} illustrate the impact of the different methods and the fact that the worst-case scenario indeed corresponds to equality of the survival functions after end of the study and no better lower-bound can be found. Here we consider continuous time since continuous parametric models are used in practice for survival extrapolation in cost-effectiveness analysis. This in particular also shows that continuity does not help in getting a better lower bound for $S_1$. 
		\begin{example}
		\label{example:surv}
		Assume that the population consists of two hidden strata defined by a latent variable $U\sim\text{Bernoulli }(0.2)$, which determines the hazard rate $\lambda_a(t|u)$ under both treatments $a=0,1$. Assume a Weibull model for the hazard rate under the control group for each stratum and hazard ratios as given below: 
		\[
		\begin{cases}
			\lambda_0(t\mid 0)=\frac{a_0}{\sigma_0}\left(\frac{t}{\sigma_0}\right)^{a_0-1}\\
			\lambda_0(t\mid 1)=\frac{a_1}{\sigma_1}\left(\frac{t}{\sigma_1}\right)^{a_1-1}\\
		\end{cases}\quad 
		\begin{cases}
			\text{HR}(t\mid 0)=\min\{0.01+4.95(t-0.8)_{+},1\}\\
			\text{HR}(t\mid 1)=\min\{0.9+0.1t,1\}\\
		\end{cases}
		\]
		with $a_0=3,$ $a_1=0.8$, $\sigma_0=0.5$, $\sigma_1=0.8$. This in particular means that the treatment is initially effective for everyone and treatment effect is waning for the individuals with $U=1$ from the beginning. For those with $U=0$, treatment efficacy starts to wane after $t=0.8$. After $t=1$ there is no treatment effect for all individuals since $\text{HR}(t|u)=1$ for $u=0,1$. {These hazard ratios are shown in the right panel of Figure~\ref{fig:surv_extrapolation}.  The resulting survival functions for the two treatment groups are shown in the left panel of Figure~\ref{fig:surv_extrapolation}}. Assume that the follow up of the study is only $\tau=0.8$, so the part behind the vertical black line in the plot is not observed.
		
		{In practice, one first extrapolates $S_0$ for example by fitting a parametric model to the observed data. To illustrate the core problem, in the left panel of Figure~\ref{fig:surv_extrapolation}, we  {ignore} estimation uncertainty and {assume} to be in the ideal world were we know the true survival functions up to $\tau$ and also $S_0$ beyond $\tau$ (dashed blue line). 
		If we would {impose waning by forcing equality of the hazard functions beyond the study duration, i.e. $\text{HR}(t)=1$ for $t>\tau$,} we would extrapolate $S_1$ using the formula $$S_1(t)=S_1(\tau)\exp\left(-\int_\tau^t\lambda_0(s)\dd s\right)=S_0(t)\frac{S_1(\tau)}{S_0(\tau)}, \quad t>\tau.$$		
		In this way}  we would get the dotted red line in the left panel of Figure~\ref{fig:surv_extrapolation}, which clearly overestimates the actual survival probabilities (dashed red line). Moreover, since the hazard ratio is not a causal comparison, forcing equality of hazards between the two
		treatment groups does not have any interpretation in terms of waning. 
		
		Sudden waning applied directly to the survival function, i.e. $S_1(t)=S_0(t)$ for $t>\tau$, is not often used in practice since it is considered a very conservative approach and  gradual waning is thought to be more plausible. 
		However, from this example we can see that, depending on how rapidly the treatment effect diminishes, the true $S_1$	may decline more sharply after $\tau$. Hence, {under the assumption that treatment is not detrimental,} considering the equality of survival curves beyond 
		 $\tau$ as a lower bound for worst-case analysis can help to quantify the uncertainty in cost-effectiveness of treatments.
	\end{example}

\begin{figure}
		\centering
		\includegraphics[width=0.5\linewidth]{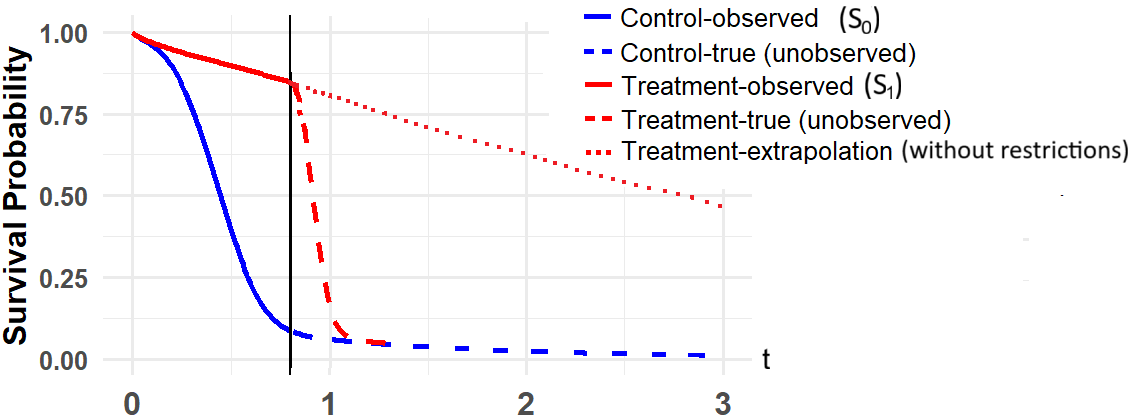}
		\caption{Survival functions for the scenario described in Example~\ref{example:surv}. Study duration is $\tau=0.8$ and the dotted red line corresponds to the extrapolated survival function for the treatment group {without any restriction related to the hazard of the control group}.} 
	\label{fig:surv_extrapolation_uncertainty}
	\end{figure}

	As seen in Figure~\ref{fig:surv_extrapolation} and also observed by other studies \citep{micallef2022msr63}, the method that is used to impose waning has a large impact on the survival function and as a result also on cost-effectiveness decisions. 
Note that the two survival functions are given by
		\[
		\begin{aligned}
S_1(t)&=\E\left[\exp\left\{-\int_0^t\lambda_0(s\,|\,U)\text{HR}(s\,|\,U)\,\dd s\right\}\right],\\
S_0(t)&=\E\left[\exp\left\{-\int_0^t\lambda_0(s\,|\,U)\,\dd s\right\}\right].
		\end{aligned}
	\]
		Under the assumption that {the protection offered by the treatment} 
		wanes at individual level stopping completely after some time $\tau$, we would have $\text{HR}(t|u)$ is increasing on $(0,\tau)$ and $\text{HR}(t|u)=1$ for $t\geq \tau$ and all $u$. As a result, for $t>\tau$ we get
	\[
	\begin{aligned}
		S_1(t)-	S_0(t)=&\,\E\left[\exp\left(-\int_\tau^t\lambda_0(s\,|\,U)\,\dd s\right)\left\{\exp\left(-\int_0^\tau\lambda_0(s\,|\,U)\text{HR}(s\,|\,U)\,\dd s\right)\right.\right.\\
		&\left.\left.\qquad\qquad\qquad\qquad\qquad\qquad\qquad-\exp\left(-\int_0^\tau\lambda_0(s\,|\,U)\,\dd s\right)\right\}\right].
		\end{aligned}
	\]
	From this, one can see that in the best-case scenario, if $\lambda_0(s|u)=0$ for $s>\tau$ and all $u$, the distance between the two survival functions remains constant after $\tau$. If $\lambda_0(s|u)$ is positive but very small, the distance between the two survival functions would decrease very slowly. On the other hand,  if $\lambda_0(s|u)$ for $s>\tau$ is very large, the distance between the two survival functions would decrease very quickly. Hence,  without further untestable assumptions, no better lower and upper bound can be obtained for $S_1(t)$, $t>\tau$ either than $[S_0(t),S_0(t)+S_1(\tau)-S_0(\tau)]$. {Without any treatment effect restriction, in practice one extrapolates $S_0$ and $S_1$ independently of each other using the fitted survival model for the corresponding group}. An illustration is shown in Figure~\ref{fig:surv_extrapolation_uncertainty}, where we again ignore the finite sample uncertainty and consider the true $S_0$ for the extrapolation.  In order to account for waning, we recommend that the best practice would be to consider the worst-case scenario of equal survival functions, i.e. $S_1(t)=S_0(t)$ for $t>\tau$. The difference between the cost-effectiveness under both scenarios, {i.e. the dotted red line and the dashed blue line in Figure~\ref{fig:surv_extrapolation_uncertainty},} would quantify the uncertainty involved and should be taken into account in decision making. 
	
%
%

	\section{Discussion}
	\label{sec:discussion}

	In this paper we discuss whether reduction of treatment efficacy over time could be detected from standard time-to-event data. The notion of waning in itself is not clearly and uniquely defined and commonly wrongly characterized as an increasing hazard ratio. {Indeed}, since the hazard ratio is not a causal comparison, it does not provide any information on waning. At first sight, it might seem that causal formulations of the hazard ratio, via principal stratification or the controlled direct effect approach, could solve the problem and allow for detection of waning in certain scenarios where the partial identification bounds are informative. In this spirit, the challenge effect was recently introduced by \cite{janvin2024quantification} to quantify waning of vaccine efficacy under an intervention that controls exposure. We show that in the simplified setting where everyone is exposed (at risk) in absence of intervention, their challenge effect is equal to the `controlled direct effect' causal hazard ratio. Moreover, we illustrate through an example that such causal hazard ratios can also provide misleading conclusions in terms of treatment effect waning. Particularly, the same survival functions can correspond to scenarios with and without waning. This means that the problem is not just about choosing the right estimand to define waning, but the observed data alone are not sufficient to distinguish between waning or no waning scenarios. Note that in this paper we restrict to data from randomized trials to show that the problem exists even under initial randomization of treatment as, in practice, there would always exist unobserved variables that affect survival at different time points.  
	
	One could try to impose further modeling assumptions, such as frailty models, to be able to detect and quantify waning. However, to do so we would need to assume that, apart from the frailty and the observed covariates, there are no further unobserved variables that affect survival at different time points.  Since this is a quite restrictive and untestable assumption, there would be no guarantee that the resulting conclusions  are not misleading. 
	
	The issue of treatment effect waning is particularly important for cost-effectiveness analysis of treatments. We conclude that there is {no correct} way to extrapolate survival functions beyond the study duration under the assumption of waning. Particularly, the common approach of forcing equality of hazards is completely arbitrary and does not have any meaning in terms of waning.  We suggest that equality of survival functions should be considered as the worst-case scenario for quantifying the existing uncertainty.

	
	
	
	


	%
 \section*{Conflict of interest}

 The authors declare that they have no conflict of interest.

	\bibliographystyle{spbasic}      
	\bibliographystyle{spphys}       
	\bibliography{references.bib}   

\end{document}